\documentclass[prx,showpacs,showkeys,preprintnumbers,amsmath,amssymb,twocolumn,superscriptaddress]{revtex4-1}
\usepackage{graphicx}
\usepackage{amsfonts}
\usepackage{color}

\def\<{\langle}
\def\>{\rangle}

\begin{document}

\title{Connecting short and long time dynamics in hard--sphere--like colloidal glasses}

\author{Raffaele Pastore}
\email{pastore@na.infn.it}
\affiliation{CNR--SPIN, Dip.to di Scienze Fisiche,
 Universit\`a di Napoli ``Federico II'', Naples,
Italy}

\author{Massimo Pica Ciamarra}
\affiliation{Division of Physics and Applied Physics, School of Physical and Mathematical Sciences, Nanyang Technological University, Singapore}

\author{Giuseppe Pesce}
\affiliation{Dipartimento di Fisica,
 Universit\`a di Napoli ``Federico II'', Naples,
Italy} 

\author{Antonio Sasso}
\affiliation{Dipartimento di Fisica,
 Universit\`a di Napoli ``Federico II'', Naples,
Italy}

\date{Received: \today / Revised version: }

\begin{abstract}
Glass--forming materials are characterized by an intermittent motion at the microscopic scale. Particles 
spend most of their time rattling within the cages formed by their neighbors, and seldom jump to a different cage.
In molecular glass formers the temperature dependence of the jump features, such as the average caging time and jump
length, characterize the relaxation processes and allow for a short--time prediction of the diffusivity. 
Here we experimentally investigate the cage--jump motion of a two--dimensional hard---sphere--like colloidal suspension,
where the volume fraction is the relevant parameter controlling the slow down of the dynamics.
We characterize the volume fraction dependence of the cage--jump features and show that,
as in molecular systems, they allow for a short time prediction of the diffusivity.
\end{abstract}

\maketitle
\section{Introduction}
The glass transition occurring in many materials can be induced by changing different control parameters. 
In molecular liquids, for example, the temperature is the relevant control parameter~\cite{Angell, Debenedetti},
while in hard sphere systems the transition is controlled by the density~\cite{Parisi_hs, Pusey, Tarzia}.
In other systems, such as attractive or soft colloids, 
both temperature and density play an important role~\cite{Trappe, Likos, de Candia}.
Despite this variety, glass--forming materials exhibit common features.
Indeed, on approaching the glass transition
one observes a dramatic increase of the relaxation time,
a vanishing diffusivity, the breakdown of the Stokes--Einstein relation~\cite{Biroli}
and the emergence of dynamical heterogeneities~\cite{DHbook}.
At the microscopic level, one observes the emergence of
an increasingly intermittent single particle motion,
both in equilibrium supercooled liquids~\cite{Intermittence, SM14} and aging glasses~\cite{Sanz}.
Indeed, in glassy systems particles spend most of their time confined within the cages formed by their neighbors, 
and seldom make a jump to a different cage.
This universality suggests that jumps might be the elementary irreversible events
allowing for the relaxation of the structural glasses~\cite{SM14}. 
If this is so, then particles move performing
a random random walk with step size 
of average length $\<\Delta r_J\>$, and average duration, $\<\Delta t_J\>$.
Since the jump duration is small with respect to relaxation time,
an important consequence of this scenario is the possibility
of determining the diffusivity $D$ on the time--scale of the jump duration,
i.e. well before the system enters the diffusive regime. Indeed,
one expects
\begin{equation}
\label{eq:diffusion}
D=\rho_J \frac{\langle\Delta r^2_J\rangle}{\langle\Delta t_J\rangle}
\end{equation}
where $\rho_J$ is the density of jumps, i.e. the fraction of particles
that are making a jump at every instant of time. We have recently 
investigated this scenario via numerical simulations
of a molecular liquid model~\cite{SM14}, through an algorithm able to segment
the trajectory of each particle in cages and jumps.
This allowed to verify that jumps are irreversible events, and 
that the relation between the features of the cage--jump motion 
and the diffusivity holds as the dynamics slow down by lowering the temperature.

In this paper we experimentally investigate whether a similar scenario holds
in hard sphere like systems, where the density is the relevant control parameter
and the temperature plays a minor role, as it simply fix the dynamical time-scale. 
This is not obvious, as the physical mechanisms responsible for the slow down
of molecular and of hard sphere systems might be different.
Indeed, in the first case the slow down occurs on cooling
as the system spends an increasing amount of time
close to minima of its potential energy landscape~\cite{Heuer03, Heuer05, Heuer12, arenzon, makse, Sciortino}. 
Conversely, in hard sphere systems the slowing
down has a purely entropic origin, and the elementary relaxation
events might not be single particle jumps, but rather structural
rearrangements involving a finite number of particles~\cite{Onuki13, Kawasaki}.
Via the experimental investigation of a two--dimensional hard--sphere--like colloidal system,
here we show that particle jumps are irreversible and that
they are short lived with respect to the relaxation time.
This allows for a short time prediction of the diffusivity of hard--sphere sphere systems
via Eq.~\ref{eq:diffusion}.

\section{Methods}
\subsection{Experiments}
\begin{figure}[t!]
\begin{center}
\includegraphics*[scale=0.26]{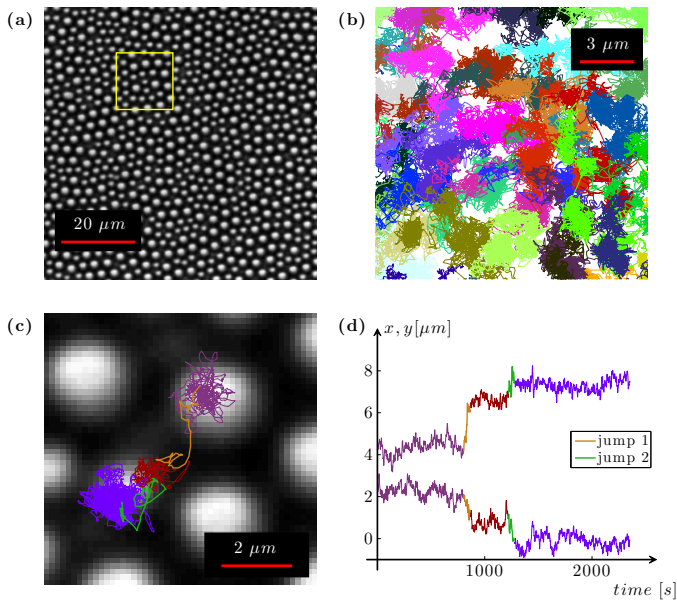}
\end{center}
\caption{\label{fig:experiment}
(a) Snapshot of the investigated system at volume fraction $\phi=0.76$.
(b) Trajectories of all particles in the region highlighted in (a).
A portion of one such trajectory segmented in cages and jumps is illustrated in real space (c)
and in the $(x,t)$ and $(y,t)$ space (d).}
\end{figure}

~
We have experimentally investigated the motion of a two--dimensional
layer of colloidal particles immersed in water.
The sample was a 50:50 binary mixture of silica beads, with 
bead diameters $3.16\pm0.08$ and $2.31\pm0.03 ~\mathrm{\mu}m$ respectively,
resulting in a $\approx 1.4$ ratio known to prevent crystallization.
The sample cell was prepared with a microscope slide and a No.1 thickness coverslip 
separated by two Parafilm stripes. Heating the whole cell up to $90\,^{\circ}C$
allowed the Parafilm stripes to melt and then to glue the two glasses. 
The resulting sample cell thickness was about 90-100 $\mathrm{\mu} m$. 
The silica particles, being heavier than water, settle on the bottom coverslip
creating a two dimensional system of free diffusing particles.
We image the system using a standard microscope equipped with a 40x objective
(Olympus UPLAPO 40XS). The images were recorded using a fast digital camera
(Prosilica GE680). At the highest volume fraction, we image roughly a thousand of particles
in the field of view of our microscope (see Fig.~\ref{fig:experiment}a). 
Particle tracking was performed using custom programs.

To avoid bacterial contamination both the bead mixture and the sample
cell were carefully washed several times with ethanol and then with distilled highly
purified MilliQ water. To avoid particle sticking through Van der Waals forces, the beads were dispersed in a water 
surfactant solution (Triton X-100, 0.2 \% v/v).
With this concentration the particles did not stick to the coverslip for days. 
The sample temperature was continuously monitored during experiments, remaining
stable within $1\,^{\circ}C$ around the room temperature ($T=22\,^{\circ}C$). 

We have investigated different volume fractions $\phi$, in the range $0.64$--$0.79$. 
At higher volume fractions the time required for the particles to settle down
in a single monolayer was too long to avoid particle sticking.

\subsection{Cage-jump detection algorithm}~
We segment each of the experimentally recorded particle
trajectory in a sequence of cages and jumps, as illustrated in Fig.~\ref{fig:experiment}c and d. 
We use an algorithm~\cite{SM14} that associates to each particle, at each time $t$,
the fluctuation of its position, $S^2(t)$, computed over the interval $[t-10t_b:t+10t_b]$, with
$t_b\simeq 1 s$ being the ballistic time. 
At time $t$, a particle is considered
in a cage if $S^2(t) < \<u^2\>$, as jumping otherwise.
Here $\<u^2\>$ is the Debye--Waller factor,
that we determine from the mean square displacement as 
in Ref.~\citenum{Leporini} and whose volume fraction dependence is 
shown in Fig.~\ref{fig:msd}b.
At each instant the algorithm gives access to the density of jumps, $\rho_J$,
defined as the fraction of particles which are jumping, and to the density of cages, $\rho_C=1-\rho_J$.
By monitoring when $S^2$ equals $\<u^2\>$, we are able to 
identify the time at which each jump (or cage) starts and ends.
That is, this approach explicitly considers that jumps are processes with a finite duration.

\section{Results}
\subsection{Glassy dynamics}
\begin{figure}[t!]
\begin{center}
\includegraphics*[scale=0.33]{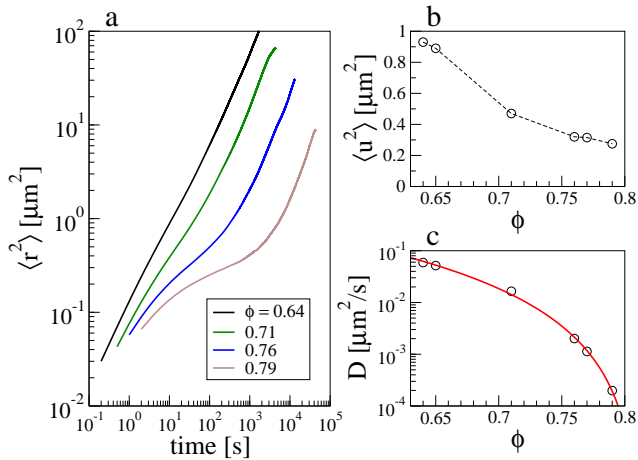}
\end{center}
\caption{\label{fig:msd}
(a) Mean square displacement for different volume fraction, as indicated.
(b) volume fraction dependence of the Debye--Waller factor. The dashed line is a guide to the eye.
(c) volume fraction dependence of the diffusivity. The full 
line corresponds to a power law fit, $D(\phi) \propto (\phi_c-\phi)^b$, with $\phi_c \simeq 0.81\pm0.01$
and $b = 2.8\pm0.02$.
}
\end{figure}

\begin{figure}[t!]
\begin{center}
\includegraphics*[scale=0.33]{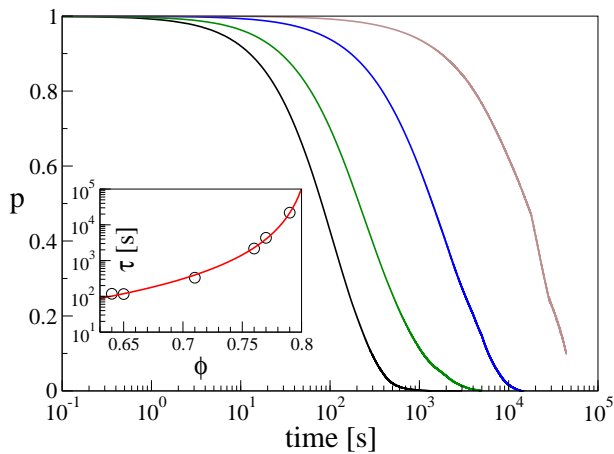}
\end{center}
\caption{\label{fig:persistence}
Persistence correlation functions for increasing values of $\phi$, from left to right.
The values of $\phi$ are as in Fig.~\ref{fig:msd}a.
The inset illustrates the volume fraction dependence of the persistence relaxation time. 
The full line is a power law fit,
$\tau(\phi) \propto (\phi_c-\phi)^{-c}$, with $\phi_c \simeq 0.81\pm0.01$
and $c = 2.6\pm0.02$.
} 
\end{figure}

We have investigated the slow dynamics of the system considering
the volume fraction dependence of the mean square displacement, $\<r^2(t)\>$,
and of the persistence correlation functions, $p(t)$, respectively
illustrated in Fig.~\ref{fig:msd}a and Fig.~\ref{fig:persistence}.
The persistence correlation function is defined, in analogy to lattice models,
as the fraction of particles that has not jumped up to time $t$~\cite{Chandler_PRE, PRL2011, Chaudhuri}.

As the volume fraction increases, the mean square displacement
develops a long plateau before entering the diffusive regime,
as usual in glass--forming systems. The value of this plateau
is the Debye--Waller factor $\<u^2\>$, and estimates the 
amplitude of the vibrational motion before the system relax.
We have measured $\<u^2\>$ as the value 
of the mean square displacement
when its logarithmic time derivative acquires the
minimum value~\cite{Leporini}. 
Fig.~\ref{fig:msd}b illustrates that $\<u^2\>$,
which is the only parameter required by the algorithm used to segment
particle trajectories, gradually decreases as the volume fraction increases. 
Fig.~\ref{fig:msd}c illustrates the volume fraction dependence
of the diffusivity $D$, that we estimate from the long time behavior $\<r^2(t)\>=Dt$.
We observe the diffusivity to decrease by three order of magnitudes
following a Mode Coupling power law behavior $D \propto (\phi_c-\phi)^b$,
with $\phi_c \simeq 0.81\pm0.01$ and $b = 2.8\pm0.02$.

Fig.~\ref{fig:persistence} shows the decay of the persistence at different volume fraction.
From this decay we have extracted the typical relaxation time, $p(\tau) = 1/e$, whose volume fraction
dependence is illustrated in the inset. 
The relaxation time is well described by a power law
functional form, $\tau(\phi) \propto (\phi_c-\phi)^{-c}$, with $\phi_c \simeq 0.81\pm0.01$
and $c = 2.6\pm0.02$. The critical volume fraction and the critical exponents describing the behavior of $\tau$
and that of $D$ are compatible. This indicates that, despite the presence
of a marked glassy dynamics, as apparent from the plateau observed
in the mean square displacement at the highest investigated volume fraction,
the system is still in the so-called mode--coupling regime.

\subsection{Cage--jump dynamics}
\begin{figure}[t!]
\begin{center}
\includegraphics*[scale=0.33]{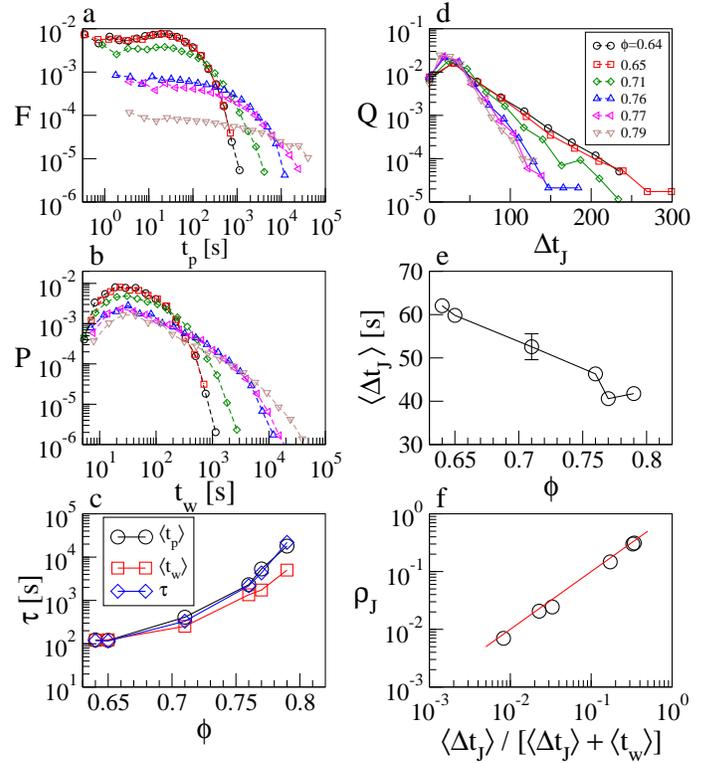}
\end{center}
\caption{\label{fig:cage_stats}
Probability distribution of the persistence time (a) and of the waiting time (b), and volume fraction
dependence of their average values (c). Probability distribution of the average jump duration (d),
and volume fraction dependence of its average value (e). Panel (f) show that the density of jumps
is fixed by the average waiting time and by the average jump duration. In panel (f) the straight line
is $y = x$; in all other panels lines are guides to the eye.}
\end{figure}

We now consider the temporal and spatial features of the cage--jump motion,
and their volume fraction dependence.
We start by considering the temporal features, summarized in Fig.~\ref{fig:cage_stats}.
Panel a illustrates the distribution $F(t_p)$ of the time
particles persist in their cage before making their first jump,
for different volume fractions. This time is measured starting
from an arbitrarily defined reference time, $t = 0$.
This distribution is of interest as directly related to the persistence correlation function,
$p(t)=1-\int_{t_p=0}^{t} {F(t_p) dt_p}$~\cite{berthier_epl2005, Hedges2007, Chandler_SE}.
Panel b illustrates the distribution $P(t_w)$ of the time particles wait in a cage
between two subsequent jumps, i.e. the cage duration. 
In the continuous time random walk
approximation~\cite{PhysRep}, these two distributions are related by $F(t_p) \propto \int_{t_p}^\infty P(t_w) dt_w$.
The two distributions are characterized by different average values, $\<t_p(\phi)\>$ and $\<t_w(\phi)\>$,
whose volume fraction dependence is illustrated Fig.~\ref{fig:cage_stats}c,
together with the volume fraction dependence of the relaxation time $\tau(\phi)$.
We observe  $\<t_p(\phi)\>$ and $\<t_w(\phi)\>$ to have a similar behaviour, and the persistence time
to scale exactly as the relaxation time, consistently with the relation between
$p(t)$ and $F(t_p)$ mentioned above. In the continuous time random walk
description of the relaxation of structural glasses, the
agreement between $\<t_p(\phi)\>$ and $\<t_w(\phi)\>$ implies the validity of the Stokes--Einstein relation,
in agreement with our system being in the mode--coupling regime.
As a further characterization of the temporal features of the cage--jump motion, 
we illustrate in Fig.~\ref{fig:cage_stats}d and e the probability distribution 
of the jump duration, $Q(\Delta t_J)$, which decays exponentially, 
and the volume fraction dependence of the average value
$\<\Delta t_J\>$, which decreases on compression.
An exponential $Q(\Delta t_J)$ distribution has been also 
observed in model molecular glasses, but in that case
the average value was found to be temperature independent~\cite{SM14}. 
Fig.s~\ref{fig:cage_stats}d,e offer us the opportunity to clarify that
the elementary process identified with a jump has a finite duration, which in the present case
can be of the order of minutes. The use of the term
`jump', which suggests the presence of short--lived events,
is only justified as the jump duration should be compared with the relaxation time
of the system. For instance, in this work the ratio $\tau/\<\Delta t_J\>$ increases
from $2$, at the smallest volume fraction with glassy features, to $\approx 250$, at the highest
volume fraction we have considered.

We finally consider that, at every instant of time, a particle is either caged or jumping.
Accordingly, by observing the system for a time of the order of $10t_b$, which
is the timescale considered by the protocol used to segment the trajectory in cages and jump,
we can measure the density of jumps $\rho_J$. This equals the probability 
that a particle is jumping at a generic time $t$, and is therefore related 
to the fraction of the total time particles spend jumping, 
\begin{equation}
\label{eq:rho_J} 
\rho_J = \frac{\langle\Delta t_J\rangle}{\langle t_w\rangle+\langle\Delta t_J\rangle}.
\end{equation}
Fig.~\ref{fig:cage_stats}f shows that this equation is verified by our data.
As mentioned above, as the dynamics slows down the jump duration becomes much smaller
than the relaxation time, so that $\Delta t_J(\phi) \ll \<t_w(\phi)\>$,
and $\rho_J \simeq \<t_w\>^{-1}$. 
We also note that in order to compute the r.h.s of the above equation one has to estimate $\<t_w\>$:
this requires to reliably sample the waiting time distribution $P(t_w)$,
an operation accomplished on a time scale of the order of the relation time $\tau$.
Conversely the l.h.s. can be estimated on a small and
density independent timescale of the order of $\< \Delta t_J\>$, 
as the only requirement is to observe a finite number of jumps.
As the density decreases, the ratio $\<\Delta t_J\>/\tau$ decreases, 
which implies that we predict a long time feature from a short time analysis.

\begin{figure}[t!]
\begin{center}
\includegraphics*[scale=0.33]{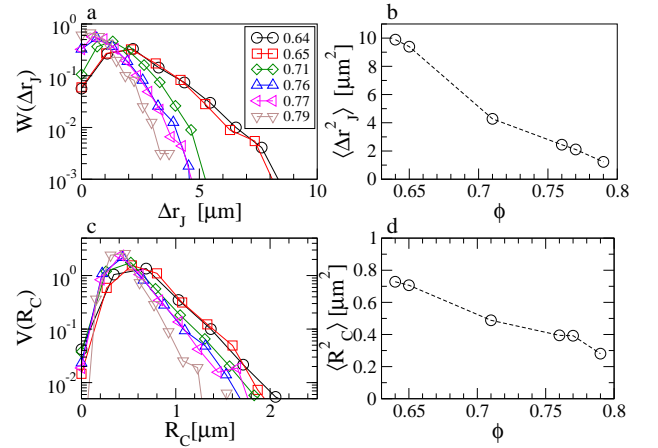}
\end{center}
\caption{\label{fig:jump_stats}
Probability distribution of the jump length (a), and volume fraction
dependence of its square average value (b). 
The same quantities are reported for the cage gyration radius in 
panel c and d respectively. }
\end{figure}

As a final characterization of the cage--jump motion, we have considered
the jump length $\Delta r_J$, defined as the distance between the centers of mass 
of two adjacent cages, and the cage gyration radius $R_C=\sqrt{\<r_i^2\>-\<r_i\>^2}$, where 
the averages run over the trajectory points $r_i$ belonging to a given cage.
The probability distribution of the jump length, $W(\Delta r_J)$, and the volume
fraction dependence of $\<\Delta r_J^2\>$  are illustrated in
Fig.s~\ref{fig:jump_stats}a and b. 
As in molecular systems~\cite{SM14}
$W(\Delta r_J)$ decays exponentially, and its average value decreases as the dynamics slow down.
We also observe the probability distribution of the gyration radius, $V(R_C)$, to decay exponentially, 
with an average value decreasing on compression, as illustrated in Fig.s~\ref{fig:jump_stats}c and d.
Fig.s~\ref{fig:jump_stats}b and d show the presence of a separation of length-scales in the dynamics,
with the average jump length exceeding the average gyration radius of the cage by at least a factor $5$ 
(at the highest investigated volume fraction).
This complements the separation of timescales observed by comparing the cage and the jump duration.



\subsection{Relating glassy and cage--jump dynamics}
The characterization of the features of the cage--jump motion
allows to verify the main point of our work, namely the possibility
of determining the macroscopic diffusivity from a short time analysis, through
Eq.~\ref{eq:diffusion}. We stress that this
relation is only valid if jumps are irreversible events, as in this case
particles behave as random walkers~\cite{SM14}, and
\begin{equation}
\label{eq:D1}
D= \lim_{t\rightarrow \infty} \frac{1}{Nt}  \sum_{p=1} ^{N} [r_p(t) - r_p(0)]^2 = \frac{1}{Nt}  \sum_{p=1} ^{N}  \theta_J^{(p)}(t)\<\Delta r^2_J\>.
\end{equation}
The last equality is obtained considering that, at time $t$,
the contribution of particle $p$ to the overall square displacement is due to
$\theta_J^{(p)}(t)$ jumps of average square size $\<\Delta r^2_J\>$.
The average number of jumps per particle,
$\<\theta_J(t)\>=\frac{1}{N} \sum_{p=1}^{N}  \theta_J^{(p)}(t)$, appearing in the last equality can be
also written as $\<\theta_J(t)\> = t/\left(\langle\Delta t_J\rangle + \<t_w \>\right)$.
Using Eq.~\ref{eq:rho_J},  $\<\theta_J(t)\>$ and $\rho_J$ can be related, $\<\theta_J(t)\>=(\rho_J /\< \Delta t_J \>) t$,
which substituted in Eq. \ref{eq:D1} finally leads to Eq.~\ref{eq:diffusion}.

\begin{figure}[t!]
\begin{center}
\includegraphics*[scale=0.33]{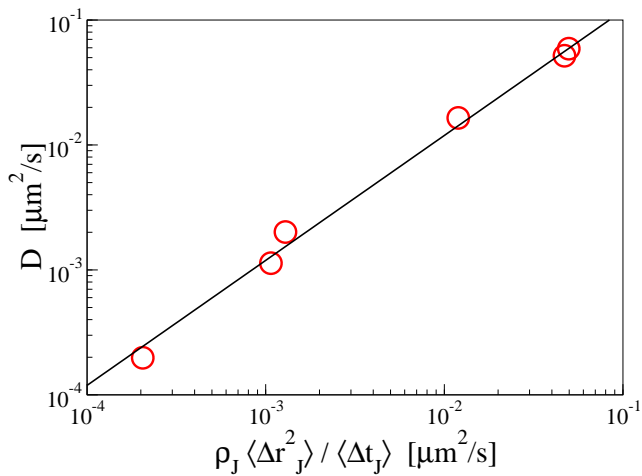}
\end{center}
\caption{\label{fig:micro_macro}
Linear dependence of the diffusion constant on  the features of the cage--jump motion.  
The solid line is the prediction of Eq.~\ref{eq:diffusion}, with
a scaling factor $\simeq 1.2$.
We stress that $D$ is estimated at long times, while $\rho_J \langle\Delta r^2_J\rangle/\langle\Delta t_J\rangle$
is estimated at short times, well before the system enters the diffusive regime. 
} 
\end{figure}

In Fig.~\ref{fig:micro_macro} we compare the measured value of the diffusivity,
with that predicted by Eq.~\ref{eq:diffusion}. We 
find $D = m \rho_J \frac{\langle\Delta r^2_J\rangle}{\langle\Delta t_J\rangle}$, 
with $m$ of the order of unity, in good agreement with the theoretical prediction.
This suggests that, at least in the investigate volume fraction
range, jumps are irreversible events.

\section{Discussion}
Our experimental investigation proves that,
in the considered volume fraction range, 
single particle jumps are the irreversible events that 
allow for the relaxation of hard sphere colloidal glasses.
This allows for a short time prediction of the diffusivity.
This result complements our earlier numerical study of a model
molecular glass~\cite{SM14}, where we also proved
single particle jumps to be irreversible events.
Indeed, we have found the same physical scenario to
capture both the slow down of the dynamics of molecular glass
formers, for which temperature is the relevant control
parameter, and of colloidal glasses, for which density
is the control parameter. 
This unifying approach is relevant considering
that alternative approaches to describe the relaxation
of molecular glasses, that identify irreversible events
as transition in the energy landscape~\cite{Heuer03, Heuer05, Heuer12},
are not relevant in hard--sphere colloidal systems.

Open questions ahead include the investigation of the validity of this
approach at higher volume fractions, where the irreversible events might involve the rearrangement
of many particles, as previously speculated~\cite{Onuki13, Kawasaki}.
In addition, it would be interesting to consider three dimensional systems,
even tough we expect the dimensionality to play a minor role,
both because structural glasses exhibit an intermittent
single particle motion in two and three dimensions, as well as because
frustration effects, that might favor collective relaxation processes, are
less relevant in high dimensions.

\bigskip
\noindent{{\bf Acknowledgement}\\
We thank A. Coniglio for discussions and
acknowledge financial support
from MIUR-FIRB RBFR081IUK.
}

\end{document}